\documentclass[11pt,a4paper]{article}

\usepackage[utf8]{inputenc} 
\usepackage[T1]{fontenc}    %
\usepackage[english]{babel} 
\usepackage[final]{microtype} 

\usepackage{amssymb,amsmath,mathtools} 
\usepackage{amsthm} 

\usepackage{xcolor,tikz,graphicx,multirow}

\usepackage{cite}
\usepackage[misc]{ifsym}

\usepackage[hmargin=0.12\paperwidth,vmargin=0.19\paperwidth,bindingoffset=0cm]%
{geometry} 

\numberwithin{equation}{section} 

\theoremstyle{plain}
  \newtheorem{thm}{Theorem}
  \newtheorem{prop}[thm]{Proposition}
  
  \newtheorem{cor}[thm]{Corollary}
\theoremstyle{definition}
  
  \newtheorem{remark}[thm]{Remark}

\title{\Large\bfseries The probability that all eigenvalues are real \\%
    for products of truncated real orthogonal random matrices}
\author{Peter J. Forrester${}^{(1)}$ and Santosh Kumar${}^{(2)}$}
\date{}

\begin{document}


\noindent
{\Large\bfseries The Probability That All Eigenvalues are Real for\\%
    Products of Truncated Real Orthogonal Random\\
     Matrices}
    
 \vspace{0.4cm}   
    
\noindent    
{\bf Peter J. Forrester${}^{1}$ and Santosh Kumar${}^{2}$}

\vspace{3em}


\noindent
{\bf Abstract}
The probability that all eigenvalues of a product of $m$ independent $N \times N$ sub-blocks of a Haar distributed random real orthogonal matrix of size $(L_i+N) \times (L_i+N)$, $(i=1,\dots,m)$ are real is calculated as a multi-dimensional integral, and as a determinant. Both involve Meijer G-functions. Evaluation formulae of the latter, based on a recursive scheme, allow it to be proved that for any $m$ and with each $L_i$ even the probability is a rational number. The formulae furthermore provide for explicit computation in small order cases.


\vspace{0.5cm}


\vspace{0.5cm}



\vspace{3cm}

\noindent
\rule{5.5cm}{0.5pt}
\vspace{0.1cm}

\noindent 
The work of PJF was supported by the Australian Research Council through Grant
DP14102613.

\noindent
\rule{5.5cm}{0.5pt}


\vspace{0.4cm}

\noindent 
{\Letter} Santosh Kumar\\
$~_{~}~~$ skumar.physics@gmail.com

\vspace{0.2cm}

\noindent 
$~_{~}~~$Peter J. Forrester\\
$~_{~}~~~$pjforr@unimelb.edu.au

\vspace{0.5cm}

\noindent 
${}^{1}$
School of Mathematics and Statistics, ARC Centre of Excellence for Mathematical and\\
$_{~}$ Statistical Frontiers, The University of Melbourne,Victoria 3010, Australia 
 
\vspace{0.3cm}

\noindent 
${}^{2}$
Department of Physics, Shiv Nadar University, Gautam Buddha Nagar, Uttar Pradesh 201314,\\
${~}_{~}$India


\newpage

\section{Introduction}
In general an $N \times N$ real matrix may have both real and complex eigenvalues. A natural question is thus to ask for the probability $p_{N,k}^{X}$ that a random real matrix $X$ chosen from a particular distribution has a specific number
$k$ of real eigenvalues. Due to the complex eigenvalues coming in complex conjugate pairs,
$k$ must have the same parity as $N$ for $p_{N,k}^{X}$ to be non-zero.

This question is of interest from a number of different viewpoints. In probability theory, for
large $N$, the distribution of $p_{N,k}^{X}$ in the case that $X=G_1^{-1} G_2$, where
$G_1, G_2$ are standard real Gaussian matrices, can be proved 
\cite{FM11} to satisfy a local
central limit theorem with
\begin{equation}\label{VM}
{\rm Var}_N \sim (2 - \sqrt{2}) \mu_N, \qquad \mu_N \asymp \sqrt{N}.
\end{equation}
The recent work~\cite{Si15} proves a central limit theorem for polynomial
 functionals of the real eigenvalues in this setting, and the relation~\eqref{VM} is found.
This latter relation was first observed in \cite{FN07} in the case of matrices $X$ drawn from the
real Ginibre ensemble, and thus with independent standard real Gaussian entries. Very recently
(\ref{VM}) has been observed in numerical simulations to hold for a wide class of random
matrix ensembles, and with a corresponding distribution consistent with a local limit theorem~\cite{GPT16}. 

The large $N$ form of the probability $p_{N,0}^X$ ($N$ even) that no eigenvalues are real
has been observed in \cite{KPTTZ15} to be intimately related to the large $s$ form of the
probability that the interval $[-s/2,s/2]$ of the real line is free of eigenvalues with the limit
$N \to \infty$ already taken. For $X$ a member of the real Ginibre ensemble, the latter
asymptotic form was computed in \cite{Fo15e}. Paper [17] also calculates the large $N$ asymptotics of $P_{N,k}^X$ for $k\ll \sqrt{N}/\log N$.
 The other extreme, the probability
$p_{N,N}^X$ that all eigenvalues are real, is known to be related to large
deviation principles \cite{GPTW16}, allowing for its asymptotic determination in the 
case of $X$ equal to the product of $m$ real Ginibre matrices \cite{Fo13}.

From the viewpoint of applications, the probability $p_{N,N}^X$ in the case $X$ of the form
$G_1^{-1} G_2$ gives the probability that random elements from a certain tensor
structure have minimal rank \cite{tB91}. With $G_1, G_2$ real Ginibre matrices
this probability can be computed exactly \cite{FM11,BF12}. The probability
$p_{N,N}^X$ with $N=2$ and $X = G_1 G_2$ was shown in \cite{La13} to have
an interpretation in quantum entanglement, quantifying when two-qubits $|\phi_1 \rangle$ and $|\phi_2 \rangle$ chosen from a
uniform distribution on the 3-sphere  are an optimal pair.

Another interest in $p_{N,k}^X$, for $X$ a real Ginibre matrix \cite{Ed97,AK07,Ma11},
the inverse of the real Ginibre matrix times  a real Ginibre matrix \cite{FM11}, and
the product of two real Ginibre matrices \cite{Fo13,Ku15,FI16}, are its special arithmetic
properties. The ability to probe these properties relies on integrable structures associated
with the computation of probabilities in these cases. In addition, the proof in the
case of  the product of two real Ginibre matrices relies on new evaluation formulae for certain Meijer G-functions   \cite{Ku15}. Integrable structures are also present in the computation
of statistical distributions relating to products of truncated real orthogonal matrices \cite{KSZ09,Fo10,IK14}. 
It is our aim in this
paper to make use of the integrable structures to give a formula for the probability 
$p_{N,N}^X$ in the case that $X$ is formed from the product of $m$ truncated
real orthogonal matrices, and to isolate arithmetic properties in the case $m=1$ (general $N$, $L_1$), the case $m=2$ ($L_1=1$, $L_2=2$, small $N$), and for $m\geq 2$ (general $N$, all $L_i$ even). The latter requires the derivation of some further evaluation formulae
for certain Meijer G-functions.
The evaluations of $p_{N,N}^X$, both in the form of a multi-dimensional integral,
and a determinant, are given in  
Section \ref{S2}. The multi-dimensional integral can be evaluated as a product of gamma
functions in the case $m=1$, making the arithmetic properties immediate. Section
\ref{S3} contains the evaluation formulae for certain Meijer G-functions in terms of
recurrences, and from this it follows that $p_{N,N}^X$ is rational when the $L_i$ 
are even. We conclude in Section \ref{S4} by implementing the recurrences in
some low order cases (equivalent to $m=1$ and $m=2$)
to give some explicit evaluations of $p_{N,N}^X$.

\section{The Probability $p_{N,N}^X$ for Products of Truncated Real Orthogonal
Matrices}\label{S2}
\subsection{Multidimensional Integral Formula}

Let $R$ be a Haar distributed random real orthogonal matrix of size $(L+N) \times (L+N)$, and
let $D$ be an $N \times N$ sub-block of $R$. A peculiarity of this setting is that
only for $L \ge N$ is the corresponding probability density
function of $D$ free of delta function constraints, and given by the
smooth function (see e.g.~\cite[Eq.~(3.113)]{Fo10})
\begin{equation}\label{1}
P(D) = {1 \over C_{N,L}} \det ( \mathbb I - D^T D)^{(L-N-1)/2},
\end{equation}
where
\begin{equation}\label{2}
C_{N,L} = \pi^{N^2/2} \prod_{j=0}^{N-1} {\Gamma((L-N+1+j)/2) \over
\Gamma((L+1+j)/2)}.
\end{equation}
Our interest is in the real eigenvalues of the product of random matrices
\begin{equation}\label{D}
P_m=D_1 D_2 \cdots D_m,
\end{equation}
where each $D_i$ is of size $N \times N$, but constructed as a sub-block of a
$(L_i + N) \times (L_i + N)$ random real orthogonal matrix. To be able to make
use of (\ref{1}) we will require each $L_i \ge N$, however the final formulae to
be obtained are well defined for all $L_i \ge 0$ and 
remain valid in this range.

\begin{prop}
Let $P_m$ be specified as in (\ref{D}). 
Define
\begin{equation}\label{W}
w_m(x) = G^{m,0}_{m,m} \Big ( {L_1/2,L_2/2,\dots,L_m/2 \atop
 0,0,\dots,0} \Big | x^2 \Big ),
 \end{equation}
where $G_{m,m}^{m,0}$ denotes a particular Meijer G-function as specified in e.g.~\cite{Lu69}.
Then for $L_i \ge 0$ ($i=1,\dots,m$),
\begin{equation}\label{pb}
p_{N,N}^{P_m} = \prod_{i=1}^m \prod_{s=0}^{N-1}
{\Gamma((L_i + 1 + s)/2) \over \Gamma((s+1)/2) }
\int_{\lambda_1 > \lambda_2 > \cdots > \lambda_N}
\prod_{l=1}^N w_m(\lambda_l) 
  \prod_{1 \le j < k \le N} (\lambda_j - \lambda_k) \,
  d \lambda_1 \cdots d \lambda_N.
\end{equation}
\end{prop}

\noindent
Proof. \quad  Following an idea in
\cite{ARRS13}, we decompose each $D_i$ so that
$
D_i = Q_i R_i Q_{i+1}^T,
$
where each $Q_i$ is real orthogonal and $R_i$ is upper triangular,
$$
R_i = \begin{bmatrix} \lambda_1^{(i)} & & & \\
&  \lambda_2^{(i)} & & \\
& & \ddots & \\
& & & \lambda_N^{(i)} \end{bmatrix} + T_i,
$$
with the matrix $T_i$ being strictly upper triangular.

To keep the working succinct, we suppose for the time being that
$L_i \ge N$ ($i=1,\dots,m$), in which each $D_i$ has probability
density function~\eqref{1}. Ignoring the label $(i)$ for the time being, we know from workings in
\cite{Ma11} that
\begin{align*}
& 
{1 \over C_{N,L}}   \int \det ( \mathbb I - D^T D)^{(L-N-1)/2} \, (dT) (dQ)\\
& \qquad = {1 \over C_{N,L}} {\pi^{N(N+1)/4} \over \prod_{j=1}^N \Gamma(j/2)}
\prod_{s=1}^N \pi^{(N-s)/2}
{\Gamma((L-N+s)/2) \over \Gamma(L/2)} \,
\prod_{s=1}^N (1 - \lambda_s^2)^{L/2 - 1} \\
& \qquad = \prod_{s=0}^{N-1} {\Gamma((L+1+s)/2) \over \Gamma((s+1)/2) \Gamma(L/2)}
\, \prod_{s=1}^N (1 - \lambda_s^2)^{L/2 - 1} .
\end{align*}
We now re-instate the $(i)$ by writing $L \mapsto L_i$, $\lambda_s \mapsto \lambda_s^{(i)}$,
and we require that
$
\lambda_s = \prod_{i=1}^m \lambda_s^{(i)},
$
where $\{\lambda_s \}$ are the eigenvalues of the product ($P_m$). In changing variables to
$\{\lambda_s \}$, and integrating out over $(dQ)$ and $(dT)$, we obtain for the joint
probability density of these eigenvalues the functional form
\begin{multline}\label{F1}
\prod_{i=1}^m \prod_{s=0}^{N-1}
{\Gamma((L_i + 1 + s)/2) \over \Gamma((s+1)/2) \Gamma(L_i/2)}  \\
\times \prod_{l=1}^N \int_{-1}^1 d \lambda_l^{(1)} \cdots
 \int_{-1}^1 d \lambda_l^{(m)}  \,
 \delta \Big (\lambda_l - \prod_{i=1}^m \lambda_l^{(i)} \Big )
 \prod_{i=1}^m  (1 - (\lambda_l^{(i)})^2 )^{L_i/2 - 1}
 \prod_{1 \le j < k \le N} (\lambda_j - \lambda_k).
 \end{multline}
 Now write
 \begin{equation}\label{F2}
 F(\lambda) = \int_{-1}^1 d \lambda^{(1)} \cdots
 \int_{-1}^1 d \lambda^{(m)}  \,
 \delta \Big (\lambda - \prod_{i=1}^m \lambda^{(i)} \Big )   \prod_{i=1}^m (1 - (\lambda^{(i)})^2 )^{L_i/2 - 1}.
 \end{equation}
 Taking the Mellin transform, we then have that
 $$
 \int_{-\infty}^\infty F(\lambda) |\lambda|^{p-1} \, d \lambda =
 \int_{-1}^1 \cdots \int_{-1}^1 \prod_{i=1}^m | \lambda^{(i)}|^{p-1}
 (1 - (\lambda^{(i)})^2)^{L_i/2 - 1} \, d \lambda^{(i)}.
 $$
 After a change of variables, these are all Euler beta integrals, and so
 $$
 \int_0^\infty F(\lambda) \lambda^{p-1} \, d \lambda = \phi(p), \qquad
 \phi(p) = {1 \over 2} \prod_{i=1}^m {\Gamma(p/2) \Gamma(L_i/2) \over
 \Gamma((p+L_i)/2)}.
 $$
Taking the inverse Mellin transform, we thus have that
 \begin{align*}
 F(x)  = {1 \over 2 \pi i}
 \int_{c-i \infty}^\infty x^{-2s} \prod_{l=1}^m {\Gamma(s) \Gamma(L_l/2) \over
 \Gamma(s+L_l/2)} \, ds 
 = \prod_{l=1}^m \Gamma(L_l/2) \,
 w_m(x),
 \end{align*}
 where $w_m(x)$ is given by (\ref{W}).
 
 Recalling the definition of $F(x)$ as given by (\ref{F2}), and substituting in
 (\ref{F1}), we obtain (\ref{pb}), although with the restriction $L_i \ge N$. This restriction was imposed because of the essential
use of the density~\eqref{1}. However, at the expense of more complex and lengthy working, it is
possible to do without (\ref{1}), making use instead of the fact that with $D$ the top $N  \times N$ sub-block of the
 $(L + N) \times (L+N)$ Haar distributed real orthogonal matrix, and $B$ the bottom $L \times
 L$ sub-block, is proportional to $\delta(D^TD + B^T B - \mathbb I_{N+L})$. In the case
 $m=1$, the necessary working is sketched in \cite{KSZ09}, with the full details, including
 a generalisation to so called induced ensembles \cite{FBKSZ12}, given in the thesis
 \cite{Fi12}. This has the consequence of allowing  (\ref{pb}) to be established for all
 $L_i \ge 0$.
 
 {}~\hfill $\square$

 In the special case $m=1$ we have
 \begin{equation}\label{Ga}
  G^{1,0}_{1,1} \Big ( {L_1/2 \atop 0} \Big | x^2 \Big ) = {1 \over\Gamma(L_1/2)} (1 - x^2)^{L_1/2-1} \chi_{|x| < 1}.
\end{equation}
Substituting in~\eqref{pb}, the resulting multidimensional integral can be
evaluated in terms of a product of gamma functions, allowing the arithmetic
properties of $p_{N,N}^{P_1}$ to be specified.

\begin{cor}\label{cor3}
We have
 \begin{equation}\label{F3}
 p_{N,N}^{P_1} = \prod_{j=0}^{N-1} {\Gamma(L_1+j) \Gamma((L_1+j)/2) \over
  \Gamma(L_1 + (N+j-1)/2) \Gamma(L_1/2)}.
  \end{equation} 
  As a consequence, for $L_1$ even, $p_{N,N}^{P_1} $ is rational, while for
 $L_1$ odd,  it is equal to a rational number times $1/\pi^{\lfloor N/2 \rfloor}$.
 \end{cor}
 
 \noindent
Proof. \quad
 Substituting (\ref{Ga}) in (\ref{pb}) gives
 \begin{align*}
p_{N,N}^{P_1} & =  \prod_{s=0}^{N-1}
{\Gamma((L_1 + 1 + s)/2) \over \Gamma((s+1)/2) \Gamma(L_1/2)} 
\int_{\lambda_1 > \lambda_2 > \cdots > \lambda_N}
\prod_{l=1}^N  (1 - \lambda_l^2)^{L_1/2-1}
  \prod_{1 \le j < k \le N} (\lambda_j - \lambda_k) \,
  d \lambda_1 \cdots d \lambda_N \\
 & =  {1 \over N!} \prod_{s=0}^{N-1}
{\Gamma((L_1 + 1 + s)/2) \over \Gamma((s+1)/2) \Gamma(L_1/2)} 
\int_{-1}^1 \cdots \int_{-1}^1
\prod_{l=1}^N  (1 - \lambda_l^2)^{L_1/2-1}
  \prod_{1 \le j < k \le N} |\lambda_j - \lambda_k| \,
  d \lambda_1 \cdots d \lambda_N. 
  \end{align*} 
  Changing variables in the integral shows that it is equal to
  $$
  2^{N L_1} 2^{N(N-3)/2} \int_0^1 \cdots \int_0^1 \prod_{j=1}^N x_j^{L_1/2-1}(1 - x_j)^{L_1/2-1}
  \prod_{1 \le j < k \le N} |x_k - x_j| \, dx_1 \cdots dx_N.
  $$
  This is a particular Selberg integral (see e.g.~\cite[Ch.~4]{Fo10}) and so can be evaluated to give
  $$
 2^{N L_1} 2^{N(N-3)/2}  \prod_{j=0}^{N-1} {(\Gamma(L_1/2+j/2))^2 \Gamma(1+(j+1)/2) \over
  \Gamma (L_1 + (N+j-1)/2) \Gamma(3/2)}.
  $$
  Substituting and simplifying, (\ref{F3}) results.
   \hfill $\square$
   
   \begin{remark}\label{Rb} 
   Scaling the matrix $D$ in (\ref{1}) by $1/\sqrt{L}$ and taking the limit $L \to \infty$
   shows that the distribution $P(D)$ tends to a Gaussian, now being proportional to
   $e^{-{1 \over 2} {\rm Tr} \, X^2}$. It has been known for some time that the probability
   of all eigenvalues being real for a real standard Gaussian matrix is equal to
   $2^{-N(N-1)/4}$ \cite{Ed97}. Indeed taking the limit $L_1 \to \infty$ in
   (\ref{F3}) reclaims this value.
   \end{remark}
 
 \subsection{Determinant formula}
 
 It is a standard exercise in random matrix theory to 
 write the multidimensional integral in terms of a Pfaffian, using a method based on integration over alternate variables due to de Bruijn~\cite{deB55}. The details depend on
 the parity of $N$. As noted in earlier studies relating to the computation of
 probabilities relating to real eigenvalues for certain random matrix
 ensembles \cite{FN07,FN08p,FM11,Fo13,FI16}, 
 the fact that the resulting matrix entries vanish in a checkerboard fashion allows the
Pfaffian, when expressed as the square root of an anti-symmetric matrix,
to be then expressed as a determinant of half the size.

\begin{prop}
With $w_m(x)$ given by~\eqref{W}, define
\begin{align}\label{alpha}
\nonumber 
&\alpha_{j,k}  = \int_{-1}^1 dx \int_{-1}^1 dy \, w_m(x) w_m(y) x^{j-1} y^{k-1} {\rm sgn} (y - x),\\
&\nu_j  = \int_{-1}^1 w_m(x) x^{j-1} \, dx.
\end{align}
We then have
\begin{equation}\label{det}
 p_{N,N}^{P_m} = \prod_{i=1}^m \prod_{s=0}^{N-1}
{\Gamma((L_i + 1 + s)/2) \over \Gamma((s+1)/2) } \det A,
\end{equation}
where for $N$ even
\begin{equation}
A = [\alpha_{2j-1,2k} ]_{j,k=1,\dots,N/2},
\end{equation}
while for $N$ odd
\begin{equation}
A =  \Big [ [\alpha_{2j-1,2k} ]_{j =1,\dots,(N+1)/2 \atop k=1,\dots,(N-1)/2} \:
[\nu_{2j-1} ]_{j=1,\dots,(N+1)/2} \Big ].
\end{equation}
Moreover, the matrix elements~\eqref{alpha} permit the evaluations
\begin{equation}\label{alp}
\alpha_{2j-1,2k} 
  = G^{m+1,m}_{2m+1,2m+1} \Big ( {3/2-j,\dots,3/2-j ;1,L_1/2+k,L_2/2+k,\dots,L_m/2+k \atop
 0,k,\dots,k;3/2-j-L_1/2,\dots,3/2-j-L_m/2} \Big | 1 \Big ), 
\end{equation}
 as well as
\begin{equation}\label{nu}
\nu_{2j-1}
 = \prod_{l=1}^m {\Gamma(j-1/2) \over \Gamma(L_l/2+j-1/2)}.
\end{equation}
\end{prop}  

\noindent 
Proof. \quad In addition to the original paper~\cite{deB55}, the de Bruijn formulae for the expression of
the multiple integral~\eqref{pb} as a Pfaffian are given in e.g.~\cite[Prop.~6.3.4 ($N$ even), Exercises 6.3 q.1 ($N$ odd)]{Fo10}.
Explicitly, for $N$ even this gives~\eqref{det} with $\det A$ replaced by ${\rm Pf} \,[\alpha_{j,k}]_{j,k=1,\dots,N}$.
The fact that $w_m(x)$ is even reveals from the definition~\eqref{alpha} that
$$
\alpha_{2j,2k} = \alpha_{2j-1,2k-1} = 0 \qquad (j,k=1,\dots,N/2),
$$
so every alternate element in the matrix $[\alpha_{j,k}$ is zero. Interchanging rows and columns so that the
zero elements are all in the top left and bottom right block and noting $\alpha_{2k,2j-1} = - \alpha_{2j-1,2k}$
shows that ${\rm  Pf} \,[\alpha_{j,k}]_{j,k=1,\dots,N} = \det A$ as required. The $N$ odd case is similar.
The evaluations~\eqref{alp} and~\eqref{nu} follow from standard Meijer G-function formulae (see e.g.~\cite{Lu69}).\\
 ${}_{}$ \hfill $\square$

\begin{remark} \label{Rc} 
Taking the limits $L_1,\dots,L_m \to \infty$, it follows from the definition
of the Meijer G-function as a contour integral \cite{Lu69} that
$$
 p_{N,N}^{P_m} \to \prod_{s=0}^{N-1}
 \Big ( {1 \over \Gamma((s+1)/2)} \Big )^m
 \left \{
 \begin{array}{ll} \det [ \tilde{\alpha}_{2j-1,2k} ]_{j,k=1,\dots,N/2}, & N \: {\rm even} \\[.2cm]
 \det \Big [ [\tilde{\alpha}_{2j-1,2k} ]_{j =1,\dots,(N+1)/2 \atop k=1,\dots,(N-1)/2} \:
[\tilde{\nu}_{2j-1} ]_{j=1,\dots,(N+1)/2} \Big ],& N \: {\rm odd} \end{array} \right.
$$
with
$$
\tilde{\alpha}_{j,k}  =
G^{m+1,m}_{m+1,m+1} \Big ( {3/2-j,\dots,3/2-j ;1\atop
 0, k,\dots, k} \Big | 1 \Big ),
\quad
\tilde{\nu}_{2j-1}  =  (\Gamma(j-1/2))^m.
$$
In keeping with Remark \ref{Rb}, this is the functional form derived in \cite{Fo13} for
the probability that all eigenvalues are real for a product of $m$ real standard Gaussian
matrices of size $N \times N$.
\end{remark}

\begin{remark} \label{Rd}
Using the working used to derive \cite[Prop.~6]{Fo13}, we can show that,
assuming $L_i \ne 0$ for all $i$,
$$
\lim_{m \to \infty}
{\prod_{i=1}^m \Gamma(L_i+2+j-1/2) \Gamma(L_i/2+k) \over
(\Gamma(j-1/2) \Gamma(k))^m} \, \alpha_{2j-1,2k} =
\left \{ \begin{array}{ll} 1, & j \le k \\
0, & j > k \end{array} \right.
$$
and thus $\lim_{m \to \infty}  p_{N,N}^{P_m}  = 1$, in accordance with an effect
first noted in \cite{La13}.
\end{remark}

\section{Evaluation of some Meijer G-functions}\label{S3}

\begin{prop}
Consider, for positive integers $\mu, \nu, j,k$, the expression
\begin{align} \label{Kmunu}
K^{\mu,\nu}_{j,k}:=\frac{\Gamma \left(j-1/2\right)}{\Gamma (\mu) \Gamma \left(\mu+\nu+j+k -3/2\right)} \sum _{r=1}^{\nu} \frac{\Gamma \left(j+k+r-3/2\right) \Gamma (\mu+\nu-r)}{\Gamma \left(j+r-1/2\right) \Gamma (\nu-r+1)}.
\end{align}
 Then we have the following finite-sum results:
\begin{align}\label{Gmunu}
&G^{m+1,m}_{2m+1,2m+1} \Big ( {3/2-j,\dots,3/2-j ;1,\mu+k,k\dots,k \atop
 0,k,\dots,k;3/2-j-\nu,3/2-j,\dots,3/2-j} \Big | 1 \Big )\nonumber\\
 &=G^{2,1}_{3,3} \Big ( {3/2-j ;1,\mu+k \atop
 0,k;3/2-j-\nu} \Big | 1 \Big)=K^{\mu,\nu}_{j,k} ,
\end{align}
\begin{align}\label{Gmunu2}
\nonumber
&G^{m+1,m}_{2m+1,2m+1} \Big ( {3/2-j,\dots,3/2-j ;1,\mu+k,\nu+k,k,\dots,k \atop 0,k,\dots,k;3/2-j-\mu,3/2-j-\nu,3/2-j,\dots,3/2-j} \Big | 1 \Big )\nonumber\\
 =\,&G^{3,2}_{5,5} \Big ( {3/2-j,3/2-j ;1,\mu+k,\nu+k \atop 0,k,k;3/2-j-\mu,3/2-j-\nu} \Big | 1 \Big)\nonumber\\
 \nonumber
 =&\sum_{\xi=1}^\mu\sum_{\eta=1}^\nu \frac{\Gamma(2\mu-\xi)\Gamma(\xi+j+k-3/2)\Gamma(2\nu-\eta)\Gamma(\eta+j+k-3/2)}{\Gamma(\mu)\Gamma(\mu-\xi+1)\Gamma(2\mu+j+k-3/2)\Gamma(\nu)\Gamma(\nu-\eta+1)\Gamma(2\nu+j+k-3/2)}\\
&\times\bigg(K_{j,k}^{\xi,\eta}+K_{j,k}^{\eta,\xi}+\frac{\Gamma^2(j-1/2)}{\Gamma(\xi+j-1/2)\Gamma(\eta+j-1/2)}\bigg).
\end{align}
We note that these Meijer G-functions are rational numbers. These results will be used in Section~\ref{S4} to obtain explicit evaluation of the probability $p^{P_m}_{N,N}$ for $m=2$ and $L_1,L_2$ even.

\end{prop}

 \noindent
Proof. \quad
We begin with the following three-term recurrence relation which is satisfied by Meijer G-functions:
\begin{align}
\label{recur}
\nonumber
&G^{m,n}_{p,q} \Big(
{a_1, ...\,a_n; a_{n+1},...,a_{p-1},a_p-1 \atop b_1, ...\,b_m; b_{m+1},...,b_q}\,\Big|z \Big)
+G^{m,n}_{p,q} \Big(
{a_1, ...\,a_n; a_{n+1},...,a_{p-1},a_p \atop b_1, ...\,b_m; b_{m+1},...,b_{q-1},b_q+1 }\,\Big|z \Big)\\
&=(a_p-b_q-1)G^{m,n}_{p,q} \Big(
{a_1, ...\,a_n; a_{n+1},...,a_p \atop b_1, ...\,b_m; b_{m+1},...,b_q},\Big|z \Big);~~~~~~~~~~~~n<p, m<q.
\end{align} 
With the aid of this relation we can construct the diagram as shown in Fig.~1.
\begin{figure}[ht!]
\centering
\includegraphics[width=0.8\linewidth]{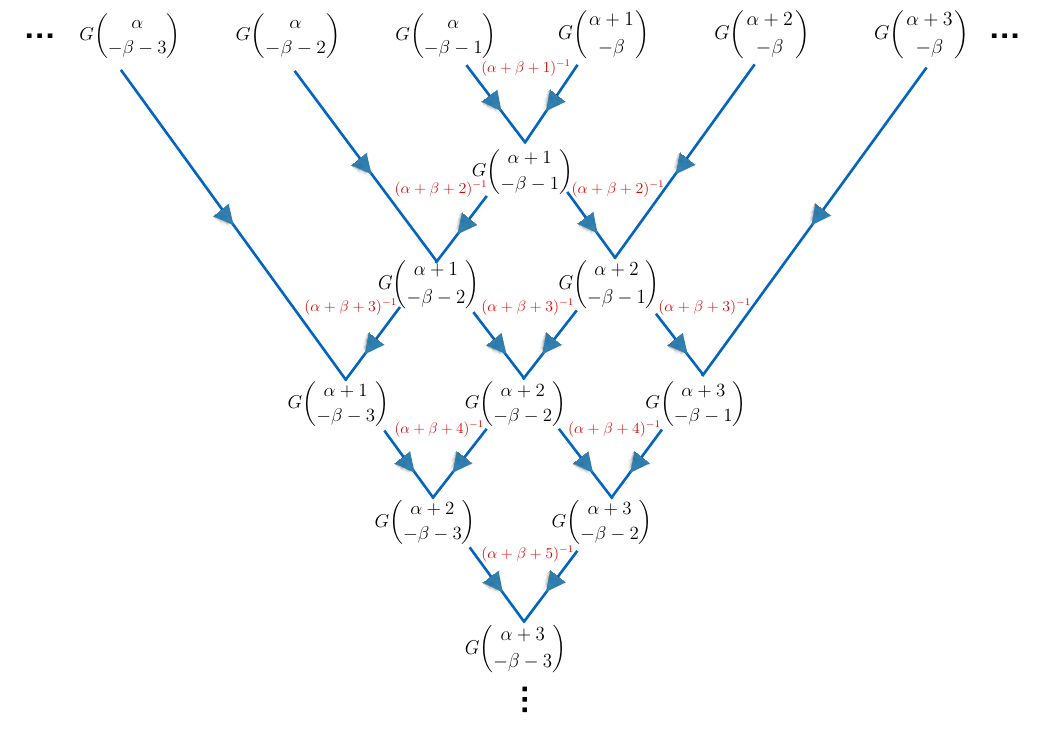}
\label{MeijerDiagram1.pdf}
\caption{Diagram facilitating the derivation of recurrence relation~\eqref{Grec}.}
\end{figure}
In this diagram we use the shorthand notation $G\binom{\alpha}{\beta}$ to represent the Meijer G-function $G^{m+1,m}_{2m+1,2m+1} \Big(
{a_1, ...\,a_m; a_{m+1},...,a_{2m},\alpha \atop b_1, ...\,b_{m+1}; b_{m+2},...,b_{2m},\beta},\Big|1\Big)$. The arrows show how successively new Meijer G-functions can be constructed from the previous ones. The `inverse terms' (in red -- colour-on-line) are the weights that have to be considered while constructing the Meijer G-functions. For example,
$G\binom{\alpha+1}{-\beta-1}=(\alpha+\beta+1)^{-1}\left[G\binom{\alpha}{-\beta-1}+G\binom{\alpha+1}{-\beta}\right],$
$G\binom{\alpha+1}{-\beta-2}=(\alpha+\beta+2)^{-1}\left[G\binom{\alpha}{-\beta-2}+G\binom{\alpha+1}{-\beta-1}\right],$
etc. All the Meijer G's below the topmost line can be constructed using those at the topmost line by following the arrows. A careful observation leads to the following recurrence relation:
\begin{align}
\label{Grec}
\nonumber
G\binom{\alpha+\mu}{-\beta-\nu}=\sum_{r=1}^{\nu} C_{\mu,\nu+1-r}\bigg(\prod_{s=r}^{\mu+\nu-1}(\alpha+\beta+s)^{-1}\bigg)G\Big({\alpha\atop -\beta-r}\Big)\\
+\sum_{r=1}^{\mu} C_{\nu,\mu+1-r}\bigg(\prod_{s=r}^{\mu+\nu-1}(\alpha+\beta+s)^{-1}\bigg)G\Big({\alpha+r\atop -\beta}\Big).
\end{align}
Here the coefficients $C_{i,j}$ are given by 
\begin{equation}\label{coeff}
C_{i,j}=\binom{i+j-2}{j-1}=\frac{\Gamma(i+j-1)}{\Gamma(i)\Gamma(j)}=\frac{1}{(i+j-1)\text{B}(i,j)}.
\end{equation}
Interestingly, these coefficients satisfy the recurrence relation
\begin{align}
C_{1,j}=1,~~
C_{i,j}=\sum_{r=1}^j C_{i-1,r},
\end{align}
and form a tilted Pascal's triangle, as depicted in Fig.~2.
\begin{figure}[ht!]
\centering
\includegraphics[width=0.25\linewidth]{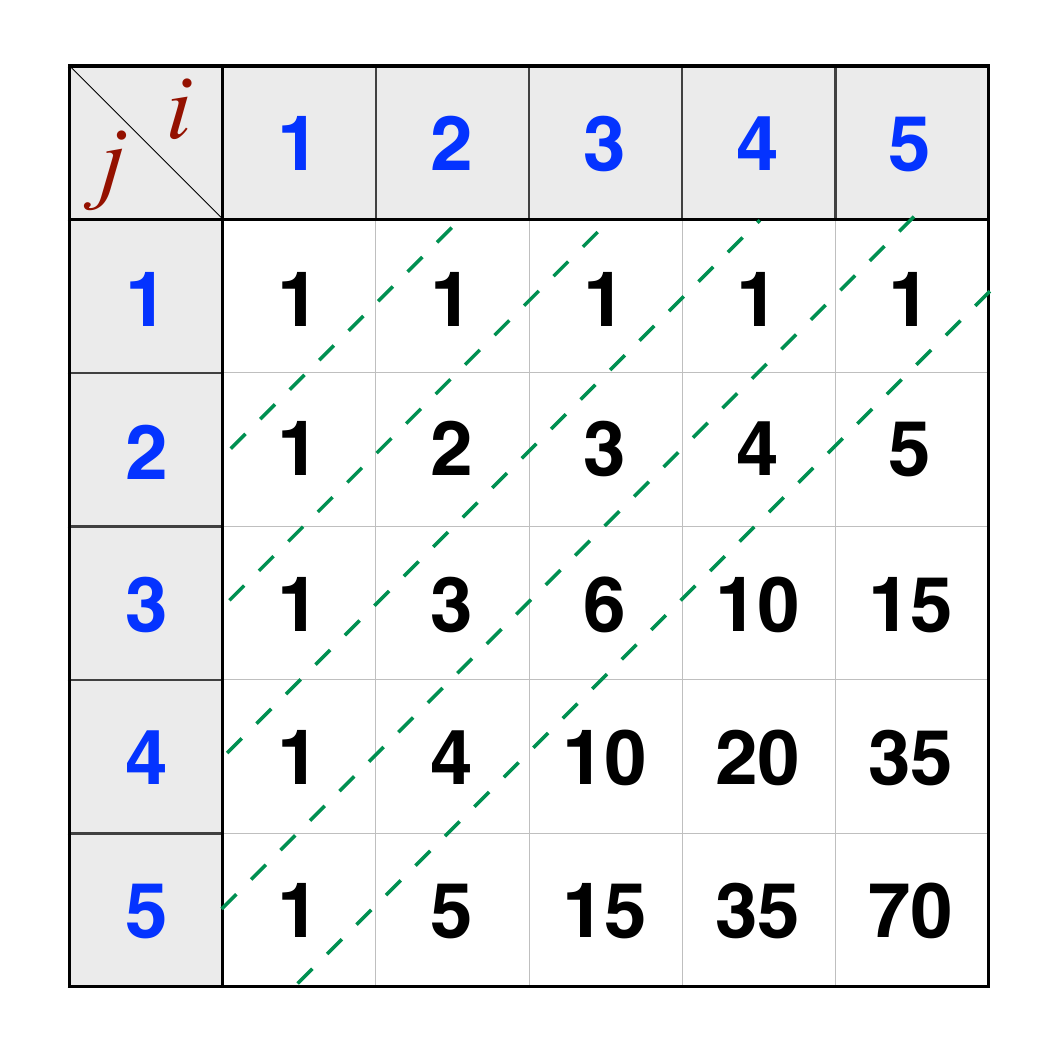}
\label{Pascal}
\caption{Some of the coefficients as defined in~\eqref{coeff}. These coefficients constitute a tilted Pascal's triangle.}
\end{figure}

Now, for non-negative integers $l_1,...,l_n$, when not all of them are 0, we have the following identities that follow from the standard contour integral formula for Meijer G-function~\cite{Lu69}:
\begin{align}
\label{id1}
G^{m+1,m}_{2m+1,2m+1} \Big({3/2-j, ...\,3/2-j; 1,l_1+k,...,l_m+k  \atop 0, k,...,k; 3/2-j,...3/2-j }\,\Big|1 \Big)=0,
\end{align} 
\begin{align}
\label{id2}
\nonumber
G^{m+1,m}_{2m+1,2m+1} \Big({3/2-j, ...\,3/2-j; 1,k,...,k  \atop 0, k,...,k; 3/2-j-l_1,...3/2-j-l_m}\,\Big|1 \Big)
=\prod_{s=1}^m\prod_{r_s=1}^{l_s} \frac{1}{(j+r_s-3/2)}\\
=\prod_{s=1}^m\frac{\Gamma(j-1/2)}{\Gamma(j+l_s-1/2)}.
\end{align} 
We note that \eqref{id2} is related to \eqref{nu}.
When $l_1=\cdots=l_m=0$, the Meijer G-function $G^{m+1,m}_{2m+1,2m+1} \Big({3/2-j, ...\,3/2-j; 1,k,...,k  \atop 0, k,...,k; 3/2-j,...3/2-j }\,\Big|z \Big)$ reduces to $G^{1,0}_{1,1} \Big({1  \atop 0}\,\Big|z \Big)$, which is a theta function involving $|z|$,
\begin{equation}\label{G1011}
G^{1,0}_{1,1} \Big({1  \atop 0}\,\Big|z \Big)=\Theta(1-|z|)=\begin{cases}1, & |z|<1,\\0, & |z|>1,\end{cases}
\end{equation}
and hence discontinuous at $z=1$. It turns out that taking its value to be $1/2$ at $z=1$ gives correct result for certain probability, as observed in Section~\ref{S4} ahead.

With the above results at our hands,~\eqref{Gmunu} follows as
\begin{align*}
&G^{m+1,m}_{2m+1,2m+1} \Big({3/2-j,\,3/2-j; 1,\mu+k,k,...,k  \atop 0, k,...,k; 3/2-j-\nu,3/2-j,...,3/2-j}\,\Big|1 \Big)
=G^{2,1}_{3,3} \Big({3/2-j; 1,\mu+k  \atop 0,k; 3/2-j-\nu}\Big|1 \Big)\\
=&\sum_{r=1}^{\nu} C_{\mu,\nu+1-r}\bigg(\prod_{s=r}^{\mu+\nu-1}(j+k+s-3/2)^{-1}\bigg)G^{2,1}_{3,3} \Big({3/2-j; 1,k  \atop 0,k; 3/2-j-r }\,\Big|1 \Big)\\
+&\sum_{r=1}^{\mu} C_{\nu,\mu+1-r}\bigg(\prod_{s=r}^{\mu+\nu-1}(j+k+s-3/2)^{-1}\bigg)G^{2,1}_{3,3} \Big({3/2-j; 1,k+r  \atop 0,k; 3/2-j}\,\Big|1 \Big)\\
=&\sum_{r=1}^{\nu}C_{\mu,\nu+1-r}\bigg(\prod_{s=r}^{\mu+\nu-1}(j+k+s-3/2)^{-1}\bigg)\bigg(\prod_{t=1}^{r}(j+t-3/2)^{-1}\bigg)+0\\
=&\frac{\Gamma \left(j-1/2\right)}{\Gamma (\mu ) \Gamma \left(j+k+\mu+\nu -3/2\right)} \sum _{r=1}^{\nu } \frac{\Gamma \left(j+k+r-3/2\right) \Gamma (\mu+\nu-r)}{\Gamma \left(j+r-1/2\right) \Gamma (\nu-r+1)}.
\end{align*}
In the second-last step we used Eqs.~\eqref{id1} and \eqref{id2}. 

Now, let us consider
\begin{align}
\nonumber
U^{\mu,\nu}_{j,k}:=
&G^{3,2}_{5,5} \Big({3/2-j,3/2-j; 1,\mu+k,k  \atop 0,k,k; 3/2-j-\mu,3/2-j-\nu }\,\Big|1 \Big)
=G^{3,2}_{5,5} \Big({3/2-j,3/2-j; 1,k,\mu+k  \atop 0,k,k; 3/2-j-\nu,3/2-j-\mu }\,\Big|1 \Big)\\
\nonumber
=&\sum_{\xi=1}^\mu C_{\mu,\mu+1-\xi}\bigg(\prod_{s=\xi}^{2\mu-1}(j+k+s-3/2)^{-1}\bigg)
\bigg(G^{3,2}_{5,5} \Big({3/2-j,3/2-j; 1,k,k  \atop 0,k,k; 3/2-j-\nu,3/2-j-\xi }\,\Big|1 \Big)\\
\nonumber
&~~~~~~~~~~~~~~~~~~~~~~~~~~~~~~~~~~~~~~~~~~~~~~~~~~~
+G^{3,2}_{5,5} \Big({3/2-j,3/2-j; 1,k,k+\xi  \atop 0,k,k; 3/2-j,3/2-j-\nu}\,\Big|1 \Big)\bigg)\\
=&\sum_{\xi=1}^\mu \frac{\Gamma(2\mu-\xi)\Gamma(\xi+j+k-3/2)}{\Gamma(\mu)\Gamma(\mu-\xi+1)\Gamma(2\mu+j+k-3/2)}\bigg(\frac{\Gamma^2(j-1/2)}{\Gamma(\nu+j-1/2)\Gamma(\xi+j-1/2)}+K^{\xi,\nu}_{j,k}\bigg),
\end{align}
\begin{align}
\nonumber
V^{\mu,\nu}_{j,k}:=
&G^{3,2}_{5,5} \Big({3/2-j,3/2-j; 1,\mu+k,\nu+k  \atop 0,k,k; 3/2-j-\mu,3/2-j }\,\Big|1 \Big)
=G^{3,2}_{5,5} \Big({3/2-j,3/2-j; 1,\nu+k,\mu+k  \atop 0,k,k; 3/2-j,3/2-j-\mu }\,\Big|1 \Big)\\
\nonumber
=&\sum_{\xi=1}^\mu C_{\mu,\mu+1-\xi}\bigg(\prod_{s=\xi}^{2\mu-1}(j+k+s-3/2)^{-1}\bigg)
\bigg(G^{3,2}_{5,5} \Big({3/2-j,3/2-j; 1,\nu+k,k  \atop 0,k,k; 3/2-j,3/2-j-\xi }\,\Big|1 \Big)+0\bigg)\\
=&\sum_{\xi=1}^\mu \frac{\Gamma(2\mu-\xi)\Gamma(\xi+j+k-3/2)}{\Gamma(\mu)\Gamma(\mu-\xi+1)\Gamma(2\mu+j+k-3/2)}\,K^{\nu,\xi}_{j,k}.
\end{align}
Finally, the result~\eqref{Gmunu2} follows as
\begin{align*}
&G^{3,2}_{5,5} \Big({3/2-j,3/2-j; 1,\mu+k,\nu+k  \atop 0,k,k; 3/2-j-\mu,3/2-j-\nu }\,\Big|1 \Big)
=G^{3,2}_{5,5} \Big({3/2-j,3/2-j; 1,\nu+k,\mu+k  \atop 0,k,k; 3/2-j-\nu,3/2-j-\mu }\,\Big|1 \Big)\\
=&\sum_{\xi=1}^\mu C_{\mu,\mu+1-\xi}\bigg(\prod_{s=\xi}^{2\mu-1}(j+k+s-3/2)^{-1}\bigg)
\bigg(G^{3,2}_{5,5} \Big({3/2-j,3/2-j; 1,\nu+k,k  \atop 0,k,k; 3/2-j-\nu,3/2-j-\xi }\,\Big|1 \bigg)\\
&~~~~~~~~~~~~~~~~~~~~~~~~~~~~~~~~~~~~~~~~~~~~~~~~~~~
+G^{3,2}_{5,5} \Big({3/2-j,3/2-j; 1,\nu+k,\xi+k  \atop 0,k,k; 3/2-j-\nu,3/2-j }\,\Big|1 \Big)\bigg)\\
=&\sum_{\xi=1}^\mu \frac{\Gamma(2\mu-\xi)\Gamma(\xi+j+k-3/2)}{\Gamma(\mu)\Gamma(\mu-\xi+1)\Gamma(2\mu+j+k-3/2)}\big(U^{\nu,\xi}_{j,k}+V^{\nu,\xi}_{j,k}\big)\\
=&\sum_{\xi=1}^\mu\sum_{\eta=1}^\nu \frac{\Gamma(2\mu-\xi)\Gamma(\xi+j+k-3/2)\Gamma(2\nu-\eta)\Gamma(\eta+j+k-3/2)}{\Gamma(\mu)\Gamma(\mu-\xi+1)\Gamma(2\mu+j+k-3/2)\Gamma(\nu)\Gamma(\nu-\eta+1)\Gamma(2\nu+j+k-3/2)}\\
&\times\left(K_{j,k}^{\xi,\eta}+K_{j,k}^{\eta,\xi}+\frac{\Gamma^2(j-1/2)}{\Gamma(\xi+j-1/2)\Gamma(\eta+j-1/2)}\right).
\end{align*}
   \hfill $\square$

\begin{remark}\label{Re}
We used the identities~\eqref{id1} and~\eqref{id2} along with the recurrence relation~\eqref{Grec} to obtain relevant Meijer G-functions for calculating $p_{N,N}^{P_m}$ for $m=1,2$ when the $L_i$ are even. It turns out that actually~\eqref{id1},~\eqref{id2} and~\eqref{Grec} can be used repeatedly in a systematic manner to obtain Meijer G-functions for any $m$ and even $L_i$. This is facilitated by the diagram shown in Fig.~3.
\begin{figure}[ht!]
\centering
\includegraphics[width=1\linewidth]{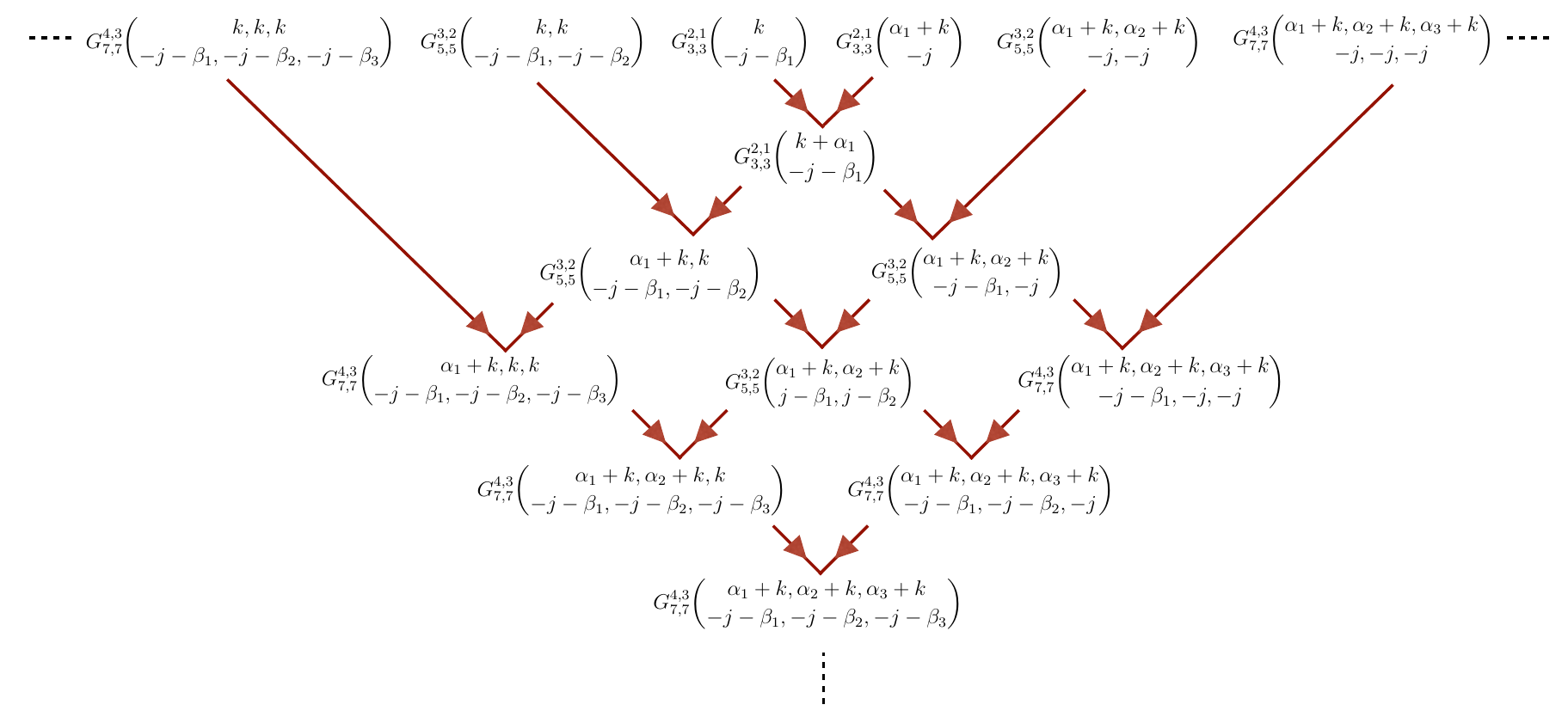}
\label{MGD2}
\caption{Diagram showing how higher order Meijer G-functions can be obtained from the lower order ones using the recurrence~\eqref{Grec}.}
\end{figure}
In this diagram we have used the notation $G_{2m+1,2m+1}^{m+1,m}\Big({a_1,...,a_m \atop b_1,...,b_m }\Big)$ to represent $G^{m+1,m}_{2m+1,2m+1} \Big({3/2-j, ...\,3/2-j;\, 1, a_1,..., a_m \atop 0, k, ...,k;\, 3/2-b_1,..., 3/2-b_m},\Big|1\Big)$. The arrows emanating from two Meijer G-functions and terminating at a third suggest the former two can be implemented in the recurrence~\eqref{Grec} to obtain the third one. The $\alpha$'s and $\beta$'s are positive integers that may serve as dummy summation variables when used in~\eqref{Grec}. The Meijer G-functions with unequal set of indices, say $G^{3,2}_{5,5}\Big({k,k\atop -j-\beta_1,-j-\beta_2}\Big)$ and $G^{2,1}_{3,3}\Big({k+\alpha_1 \atop -j-\beta_1}\Big)$ can be used in the recurrence relation by noting that the latter is same as $G^{3,2}_{5,5}\Big({k+\alpha_1,k \atop -j-\beta_1,-j}\Big)$. We also note that the Meijer G-functions at the top are already known from~\eqref{id1} and~\eqref{id2}.

Now, since the recurrence relation involves multiplication of the Meijer G's with rational coefficients, and the initial functions at the top are all rationals, it follows that the repeated application of the recurrence relation also produces rational numbers. Therefore, eventually the probability $p_{N,N}^{P_m}$ for any $m$ with each $L_i$ even is always a rational number.
\end{remark}

\begin{remark}\label{Rf}
Computer algebra, using Mathematica~\cite{Mathematica}, suggests the following results:
\begin{align}
\label{GEv1}
G_{3,3}^{2,1}\Big({3/2-j;1,\mu +1/2+k \atop 0,k;3/2-j-1/2 }\Big|1\Big)
=\frac{\Gamma (k) }{\sqrt{\pi }\, \Gamma \left(j+k+\mu -1/2\right)}\sum _{\alpha =1}^k \frac{\Gamma (\alpha +\mu ) \Gamma \left(j+k-\alpha -1/2\right)}{\Gamma \left(\alpha +\mu +1/2\right) \Gamma (k-\alpha +1)},
\end{align}
\begin{align}
\label{GEv2}
G_{3,3}^{2,1}\Big({3/2-j;1,1/2+k \atop 0,k;3/2-j-\nu-1/2 }\Big|1\Big)&=\frac{\Gamma (k)}{\Gamma \left(\nu +1/2\right) \Gamma \left(j+k+\nu -1/2\right)}
 \sum _{\alpha =1}^k \frac{ \Gamma (\alpha +\nu ) \Gamma (j+k-\alpha -1/2)}{\Gamma (\alpha +1/2) \Gamma (k-\alpha +1)}.
\end{align}
\end{remark}
It is clear that, when evaluated, these two yield $1/\pi$ times a rational number. Also, when used in a recurrence relation similar to~\eqref{Grec}, these two lead to $G_{3,3}^{2,1}\Big({3/2-j;1,\mu +1/2+k \atop 0,k;3/2-j-\nu-1/2 }\Big|1\Big)$, and hence we have for positive integer $\mu$,
\begin{align}
\label{GEv3}
\nonumber
G_{3,3}^{2,1}\Big({3/2-j;1,\mu +1/2+k \atop 0,k;3/2-j-\mu-1/2 }\Big|1\Big)=&\sum _{r=1}^{\mu } \sum _{\alpha =1}^k \frac{\Gamma (k) \Gamma (r+\alpha ) \Gamma (2 \mu -r) \Gamma \left(j+k-\alpha -1/2\right)}{\Gamma (\mu ) \Gamma (k-\alpha +1) \Gamma (\mu-r +1) \Gamma \left(j+k+2 \mu -1/2\right)}\\
&\times\bigg( \frac{1}{\sqrt{\pi }\, \Gamma \left(r+\alpha +1/2\right)}+\frac{1}{\Gamma \left(\alpha +1/2\right) \Gamma \left(r+1/2\right)}\bigg).
\end{align}
Clearly, this also equals $1/\pi$ times a rational number. This result can be used in the determinantal formula~\eqref{det} for $m=1$ and odd $L$.

It appears that 
$$G_{5,5}^{3,2}\Big({3/2-j,3/2-j;1,\mu +1/2+k,\nu +1/2+k \atop 0,k;3/2-j-\mu-1/2,3/2-j-\nu-1/2 }\Big|1\Big)~~ {\rm and} ~~ G_{5,5}^{3,2}\Big({3/2-j,3/2-j;1,\mu+k ,\nu+1/2+k \atop 0,k;3/2-j-\mu,3/2-j-\nu-1/2 }\Big|1\Big),$$
with the first being the analog of~\eqref{Gmunu2}, do not possess such simple arithmetic structures.  For instance, we find that for $m = 2, L_1 = 1, L_2 = 2$,~\eqref{W} reduces to $w_m(x) = (2/\sqrt{\pi}) \tanh^{-1}(\sqrt{1-x^2})\Theta(1-|x|^2)$, which when used in~\eqref{alp} leads to the Meijer G-values $\alpha_{1,2}=(20+8\mathcal{G})/(3\pi)$, $\alpha_{1,4}=(181+162\mathcal{G})/(90\pi)$, $\alpha_{3,2}=(17-6\mathcal{G})/(15\pi)$, and $\alpha_{3,4}=(1157+450\mathcal{G})/(3780\pi)$. Here $\mathcal{G}\approx 0.915966$ is Catalan's constant. Using these in~\eqref{det} gives the probability values as $p_{2,2}^{P_2}=(2\mathcal{G}+5)/(3\pi)$, $p_{3,3}^{P_2}=(38\mathcal{G}-1)/(30\pi)$, and $p_{4,4}^{P_2}=(29412\mathcal{G}^2+10612\mathcal{G}-6767)/(25200\pi^2)$.

\section{Some special cases}\label{S4}

\subsection*{\tikz\draw[black,fill=black] (0,0) circle (.5ex); $ \boldsymbol{L_1=\cdots=L_m=0}$}

This case corresponds to the product of $m$ orthogonal matrices each of
dimension $N$, which is again an $N$-dimensional orthogonal matrix. For the trivial case of $N=1$,~\eqref{det} gives the probability as 1, as expected. For $N=2$, with $G^{1,0}_{1,1} \big({1  \atop 0}\,\big| 1 \big):=1/2$, as discussed near~\eqref{G1011}, we get the probability value 1/2. This is in conformity with the fact that all $2\times2$ orthogonal matrices are either rotations (almost surely complex eigenvalues) or reflections
(real eigenvalues $\pm 1$) with equal probability. For $N\geq 3$, the determinant in~\eqref{det} vanishes and hence gives the probability as 0. To summarize, the probability of all eigenvalues real for the product of $m$ number of $N$-dimensional orthogonal matrices is
\begin{align*}
p_{N,N}^{P_m}=\begin{cases}
1, & N=1,\\
1/2, & N=2,\\
0, & N\geq 3.
\end{cases}
\end{align*}

\subsection*{\tikz\draw[black,fill=black] (0,0) circle (.5ex); $ \boldsymbol{ L_1>0,L_2=\cdots= L_m=0}$}

This scenario gives the probability of all eigenvalues real for product of the $N$-dimensional block of a $L_1+N$ dimensional orthogonal matrix and $m-1$ orthogonal matrices of dimension $N$. This probability turns out to be same as the probability of all eigenvalues real for the $N$-dimensional block of a $L_1+N$ dimensional orthogonal matrix, i.e., the product with other $m-1$ orthogonal matrices does not change the probability. This can be seen from~\eqref{W} that, when only one of the $L_i$ is nonzero (here $L_1$), the weight function $w_m(x)$ reduces from $G^{m,0}_{m,m} \Big({L_1/2,...,L_m/2  \atop 0,..., 0 }\,\Big|x^2 \Big)$ to $G^{1,0}_{1,1} \Big({L_1/2  \atop 0 }\,\Big|x^2 \Big)$. Thus the probability, as given by~\eqref{pb}, becomes same as that for the case $m=1$, and is obtained using~\eqref{F3}.

We also note that if we consider $L_1=2\mu>0$ then using~\eqref{Gmunu} we have
\begin{align}
\nonumber
G^{m+1,m}_{2m+1,2m+1} \Big({3/2-j,\,3/2-j; 1,\mu+k,k,...,k  \atop 0, k,...,k; 3/2-j-\mu,3/2-j,...,3/2-j}\,\Big|1 \Big)\nonumber
=G^{2,1}_{3,3} \Big({3/2-j; 1,\mu+k  \atop 0,k; 3/2-j-\mu }\,\Big|1 \Big)=K^{\mu,\mu}_{j,k},
\end{align}
where $K^{\mu,\mu}_{j,k}$ is given by~\eqref{Kmunu}.
This result can be used in the determinantal formula~\eqref{det} to calculate the probability $p_{N,N}^{P_m}$ with $L_1=2\mu,L_2=\cdots= L_m=0$. Similarly, equations~\eqref{GEv1} and~\eqref{GEv3} can be used to calculate the probability for $L_1=2\mu+1,L_2=\cdots= L_m=0$. However, these yield the same value for $p_{N,N}^{P_1}$ as given by
(\ref{F3}), as they must.

\subsection*{\tikz\draw[black,fill=black] (0,0) circle (.5ex); $ \boldsymbol{ L_{1}>0, L_{2}>0,L_3=\cdots=L_{m}=0}$}

Here we consider the case when all but two of the $L_i$ are nonzero, say $L_1$ and $L_2$. Applying reasoning similar to that in the preceding case, we find that the probability $p_{N,N}^{P_m}$ in the present scenario is the same as the probability for $m=2$, i.e., the probability $p_{N,N}^{P_2}$ of all eigenvalues real for product of $N$-dimensional blocks of orthogonal matrices of dimensions $L_1+N$ and $L_2+N$, respectively. In this case we do need to calculate the determinant and therefore need the values of Meijer G-functions. Equation~\eqref{Gmunu} can be used to obtain the explicit answers when the $L_i$ are even: $L_1=2\mu, L_2=2\nu$. As clear from the form of~\eqref{Gmunu} these probabilities are rational numbers. When one of $L_1, L_2$ is odd, or both are odd, such a simple arithmetic structure does not seem to exist, as already indicated in Remark~\ref{Rf}. 

In Table~\ref{meq2} we present probability values for various combinations of $N$ and even $L_1, L_2$, giving the explicit rational numbers.
\begin{table}[h!]
\renewcommand{\arraystretch}{1.5}
\caption{Exact values and numerical values (6 significant digits) for some probabilities $p_{N,N}^{P_2}$. }
\centering
{\small \begin{tabular}{|c|c|c|c|c| }
\hline
 \multirow{2}{*}{$N$}  & \multirow{2}{*}{$L_1$} & \multirow{2}{*}{$L_2$} & \multicolumn{2}{|c|}{$p_{N,N}^{P_2}$}   \\ \cline{4-5}
&  & & Exact  & Numerical value   \\
\hline\hline
2 (3) & 2  & 2 &$\frac{20}{27}$ ($\frac{1312}{3375}$) & $0.740741$ ($0.388741$)\\
\hline
2 (3) & 2 & 4 & $\frac{1184}{1575}$ ($\frac{4544}{11025}$)& $0.751746$  ($0.412154$)\\
\hline
2 (3)& 2 & 6 & $\frac{6112}{8085}$ ($\frac{665216}{1576575}$)& $0.755968$ ($0.421937$)\\
\hline
2 (3)& 4 & 4 & $\frac{97984}{128625}$ ($\frac{1504768}{3472875}$)& $0.761780 $  ($0.433292$)\\
\hline
2  (3)& 4 & 6 & $\frac{649984}{848925}$ ($\frac{161046016}{364188825}$)& $0.765655$ ($0.442205$)\\
\hline
\end{tabular}}
\label{meq2}
\qquad
\end{table}



\providecommand{\bysame}{\leavevmode\hbox to3em{\hrulefill}\thinspace}
\providecommand{\MR}{\relax\ifhmode\unskip\space\fi MR }
\providecommand{\MRhref}[2]{%
  \href{http://www.ams.org/mathscinet-getitem?mr=#1}{#2}
}
\providecommand{\href}[2]{#2}

\end{document}